\documentclass[12pt]{article}
\usepackage[utf8]{inputenc}

\usepackage[left=1in,right=1in,top=1in,bottom=1in]{geometry}
\usepackage{array}
\usepackage{booktabs} 
\usepackage{multirow}
\usepackage{colortbl}
\usepackage{authblk} 
\usepackage{graphicx} 
\usepackage{comment}
\usepackage{breakcites}
\usepackage[breaklinks=true]{hyperref}
\usepackage{amsmath}
\usepackage{amssymb}
\usepackage[dvipsnames]{xcolor}
\usepackage{lineno}
\usepackage{adjustbox}

\title{Prosecution of complex criminal networks: a multilevel ERGMs approach to CICIG's judicial cases}

\author[1,2]{H. Waxenecker} 
\author[3]{I. Luna-Pla$^*$}
\author[4]{J.R. Nicolás-Carlock\footnote{Correspondence: waxenecker@fss.muni.cz, issalunapla@gmail.com, jnicolas@unam.mx}}

\affil[1]{\normalsize Faculty of Social Studies, Masaryk University, Czech Republic}
\affil[2]{\normalsize Faculty of Spatial Sciences, University of Groningen, Netherlands}
\affil[3]{\normalsize Independent Analyst in Criminal Networks and Corruption}
\affil[4]{\normalsize Institute of Physics, National Autonomous University of Mexico, Mexico}
\date{}

\begin{document}

\maketitle


\begin{abstract}

Prosecutors are essential in combating organized crime, making key decisions about prosecution, target selection, and structuring imputation strategies. Despite their importance, the configuration of these strategies remains empirically underexplored. This study engages with that premise by considering the cases investigated by the International Commission Against Impunity in Guatemala (CICIG) and focusing specifically on the role of prosecutors, aiming to uncover how their discretionary decisions translated the CICIG mandate into operational practices intended to achieve systemic deterrence, and to what extent, can we talk about deterrence effectiveness. The research employs a multilevel Exponential Random Graph Model (ERGM) analysis, integrating three networks: the criminal network of actors involved in illegal activities, the legal framework network that represents offenses, and the prosecution network that connects actors to offenses. The model assesses whether the observed network aligns with deterrence-based theoretical assumptions and examines how punishment severity can be effective when it disrupts functional ties that sustain criminal activity—both through long-term sanctions and by increasing the perceived threat of punishment among co-offenders. This approach underscores the need for prosecutorial strategies to evolve beyond a case-by-case model toward a multi-case, multi-offender imputation framework that fully integrates intelligence and data-driven analysis to dismantle criminal networks.

\end{abstract}


\section{Introduction}

Prosecutors within judicial systems play a crucial role in deterring illicit networks. In recent years, national and transnational criminal organizations have become more powerful than ever, infiltrating governments, diversifying illicit businesses, expanding their territories, securing political support and recruiting multi-sector actors in various regions of the world. Under this increasing complexity, the prosecutors' decisions on whether to prosecute, whom to charge within a group of individuals, what charges to bring, and the overall strategy for presenting the case before a judge, have become paramount and, therefore, require of high degree of intelligence and innovative deterrence strategies.

The theories of organized crime have extensively examined the capacity of law enforcement to disrupt large criminal networks, such as cartels or terrorist organizations, through the analysis of actors and activities \cite{Bright2021, OKane2015}. However, a simplistic understanding of these networks often leads to counterproductive strategies, such as over-criminalizing entire groups, resorting to heavy-handed policing or military interventions, and implementing mass incarceration policies—strategies that might yield unsustainable outcomes and whose effectiveness are questionable \cite{Bottoms2004, Hazen2010}. This debate is significant because it shapes contemporary trends in punishment and legal interventions design, highlighting the need for empirical research on how prosecutorial decisions are made to strategically target individuals and organizations, with the aim of dismantling networks and deterring criminal activity \cite{Albonetti1987, Lynch2018, Barno2021}.

A modern and paradigmatic example, in which the selection of charges and prosecutorial strategies played a mayor role on the success of legal interventions for combating high-level criminality, is that of the Guatemalan cases handled by the International Commission Against Impunity in Guatemala (CICIG) and the Special Prosecutor’s Office Against Impunity (FECI) of the Public Prosecutor's Office. The CICIG is lauded for revealing the nation's power structures, which encompassed presidents and ministers, military members \cite{CICIG2019}, courts, Congress \cite{Call2021, CallandHallock2020}, political parties, and organized crime groups \cite{Beltran2020} (see Appendix A for a brief history of the CICIG). 

The CICIG cases are significant for the shift in strategy from prosecuting isolated incidents to systematically linking cases, thereby uncovering underlying criminal structures and networks. This methodological shift involved employing advanced investigative techniques, enabling the analysis of substantial data sets from varied formats and sources. Such strategies enhanced the capability for collecting robust evidence against individual suspects and facilitated the execution of group trials, leading to more thorough prosecutions (CICIG, 2019). However, while the punitive intervention might have weakened the criminal networks' economic and political influence for some time, the lack of robust analytical tools made it challenging to gauge its prosecution strategy and long-term impact \cite{Luna-Pla_2024}.

Based on CICIG's information, this research presents an innovative socio-legal network analysis of CICIG's prosecution strategy effectiveness based on criminal networks and deterrence theory. The study draws on a criminal network of 189 actors and 250 imputations involving 21 distinct legal offenses, reconstructed from eight major cases prosecuted by CICIG-FECI. The dataset is conceptualized as a multilevel structure, linking actors (criminal network), offenses (legal framework), and their imputation ties (prosecution network). Beyond treating prosecution merely as a legal mechanism for individual accountability, it is view as a systemic strategy for shaping relational ties across these interlinked levels, thereby aiming to weaken and disrupt the organizational dynamics of illicit political–economic networks.

This inferential network-based method \cite{prell2024, cranmer2020}, strongly inspired on the multilevel network analysis of socio-ecological systems \cite{bodin2016theorizing, barnes2017, barnes2022, lazega2008}, allows to theorize about preferred prosecution configurations based on specific patterns of criminal connections and the structure of the existing legal framework. Furthermore, it contributes to clarifying the role of prosecution strategies in general deterrence, offering a framework for addressing and dismantling such networks.

\section{Conceptual approach}

\subsection{Theoretical framework}

This research is grounded in the intersection between deterrence criminal theory and network science theory, particularly as it applies to the role of prosecutors in deterring illicit criminal networks. The conceptual framework integrates organized crime deterrence theory, prosecutorial discretion, and effectiveness in the organized crime and network analysis literature. The aim is to evaluate how prosecutors can, within their powers, strategically use legal imputations to disrupt resilient, diversified, and corruption-enabled criminal structures.

\subsubsection{Deterrence theory}

At the core of mainstream deterrence theory lies the assumption that rational actors weigh the costs and benefits of engaging in criminal conduct \cite{Nagin1998, Becker1968, Mendes2004}. This foundational perspective, rooted in economics and legal theory, has been explored across disciplines through both preventive and enforcement lenses. Scholars have examined why individuals commit crimes, how legal and policy mechanisms influence incentives and recidivism, and how the certainty and severity of judicial intervention shape patterns of criminality across geographic contexts and offense types \cite{Mendes2004, Nagin1998}. While recent research has broadened the deterrence framework to incorporate rehabilitation and broader social determinants of crime, punitive responses have traditionally remained central to the discourse \cite{ChalfinandMcCrary2017, Bottoms2004}.

Preventative deterrence models, employed by law enforcement and judicial systems alike, tend to focus on individual offenders using a "carrot and stick" approach. These models assume that behavior is shaped by the perceived certainty, severity, and swiftness of punishment \cite{Becker1968, ChalfinandMcCrary2017, Ariel2020, Nagin1998}. Rational choice theory underpins this strategy, positing that deterrence operates on two levels: specific deterrence, which targets individual behavioral change, and general deterrence, which seeks to influence group or societal norms through the threat of punishment \cite{Cook1980}. The effectiveness of both types of deterrence has sparked extensive debate, particularly regarding which component—certainty, severity, or celerity—exerts the greatest impact, and how these elements interact within broader judicial and enforcement systems.

We align with recent studies in emphasizing the frequently underexplored role of prosecutors selecting charges with the practical application of deterrence theory \cite{Bielen2024, Barno2021}. Although empirical studies have traditionally focused on policing and sentencing outcomes \cite{Cook1980, Braga2019}, prosecutorial discretion is equally critical. Prosecutors decide who is charged, what charges are brought, how cases are framed \cite{bellin2020, Colvin2019}, and which offenses are prioritized \cite{Albonetti1987, Lynch2018, Barno2021}. Although prosecutors occupy only one stage in the broader judicial process (their decisions are subject to judicial review and sentencing), they shape not only the legal trajectory of a case but also the public’s perception of the likelihood of enforcement—an element consistently found to be more impactful in deterring crime than severity of punishment alone, particularly in environments plagued by impunity \cite{Becker1968, Nagin2013, Bielen2024}.

The Guatemalan case, specifically the joint prosecutorial strategy implemented by CICIG-FECI \cite{Michel2021}, exemplifies a group-based deterrence model. Their approach aimed not only to increase the perceived threat of punishment for individuals within criminal networks but also to disrupt the organizational structure of those networks themselves. Unlike traditional deterrence strategies that emphasize prevention and policing programs \cite{Braga2019, Denley2025}, the CICIG-FECI model sought to disable key actors through coordinated, multi-case imputations and legal interventions, an approach that remains under-theorized in conventional legal and criminological literature.

To the effect, deterrence theory offers limited insight into how prosecutorial actions influence collective behavior or disrupt interconnected criminal networks. First, it is primarily designed to explain the behavior of individual offenders acting alone \cite{Nagin1998}. Second, while deterrence theory focuses on preventing crime, prosecutors operate at a later stage in the judicial process, aiming to disrupt ongoing criminal activity. Third, criminological and economic research has traditionally emphasized the relationship between crime rates and the threat of punishment from the judicial system \cite{Cook1980}, with insufficient attention to how group dynamics, co-offending structures, and organizational resilience sustain illicit activities over time \cite{Kleemans2012}.

To address this gap, we propose a network-based framework for understanding systemic or group deterrence. This approach enables the analysis of prosecutorial decisions as strategically coordinated interventions across multiple cases and actors. Rather than viewing prosecution solely as a legal mechanism for individual accountability, we conceptualize it as a systemic tool for weakening the relational ties of criminal networks, ultimately aiming to influence their dynamics, cohesion, and capacity for reorganization.

\subsubsection{Criminal networks theory}

Network science offers an increasingly relevant lens through which to understand deterrence in the context of organized crime \cite{KrajewskiDellaPosta2022, Bouchard2020, Bright2021}. From a criminal networks perspective —which encompasses criminal and non-criminal individuals, organized crime groups, and complicit state actors \cite{Morselli2009, Carrington2016, GimenezSalinas2014}— deterrence literature addresses questions related targeting individuals or nodes within a network with the ultimate goal of controlling, dismantling, or inhabilitating \cite{Sparrow1991, Morselli2009}. Criminal networks studies also address the elements of resilience and self organization after punitive and non punitive interventions, recidivism of offenders, and network structure metrics \cite{Morselli2016, daCunha2018, Bouchard2020}. Ultimately, this conceptual framework bridges normative legal theory with empirical analysis, enabling a richer understanding of how justice systems can disrupt the regenerative and adaptive capacity of criminal networks, particularly those bolstered by corruption \cite{luna2020corruption, Martins2022}.

While this framework acknowledges that targeting central or influential actors may enhance deterrent effects, this study does not assess node centrality or assign weights to individual actors. Although prosecutorial decisions often reflect considerations of actor visibility and political influence, the exclusion of formal centrality metrics is a deliberate methodological choice. Rather than evaluating whether criminal networks were effectively dismantled, our objective is to understand how CICIG’s institutional mandate to combat corruption and disrupt illicit networks was operationalized as a prosecutorial deterrence strategy. Accordingly, we focus on structural patterns and prosecutorial configurations that span multiple cases and offenses—particularly in contexts involving co-offenders (individuals collaborating within the same case) \cite{Bouchard2020} and multi-offenders (individuals charged with multiple offenses across different cases) \cite{asp2019, audenaert2021}.

From this perspective, criminal activity is situated within a broader universe of networked interactions that also encompasses policy coalitions, joint ventures, business partnerships, political alliances, and elite social circles. In such environments, corruption—understood as a crime of the powerful, facilitated through occupational roles for economic or political gain \cite{Joseph2021} —frequently acts as a catalyst, enabling overlapping forms of criminal conduct and reinforcing structural ties between public and private sector actors \cite{vonLampe2006, Albanese2018, Joseph2021}. A central distinction between criminal network research and studies on peer influence or neighborhood effects is that criminal network analysis focuses on actors already engaged in illicit activity. As such, the core inquiry shifts away from the etiology of crime and toward understanding its organization, structure, and systemic consequences \cite{Carrington2016, GimenezSalinas2014, vonLampe2006}.

By analyzing prosecutorial strategies through a multilevel network lens —grounded in Guatemala’s group imputation model— this study aims to advance deterrence theory by exploring how prosecutors, and ultimately courts, may generate environments where legal enforcement is perceived as both swift and certain. This is particularly relevant in contexts like Guatemala, where impunity has historically undermined the rule of law.

\subsubsection{A network-based approach to prosecutorial effectiveness}

In this study, we seek to understand how prosecutors can effectively target co-offenders, multi-offenders, and corruption-related behaviors within complex criminal networks. Evaluating prosecutorial effectiveness—particularly in terms of prevention and deterrence has long been a challenge in criminology. A key limitation lies in the scarcity of rigorous impact evaluations, largely due to restricted access to empirical data. This constraint is especially acute in cases involving organized crime, where the operations of law enforcement and prosecutorial agencies are often confidential and not readily accessible to researchers \cite{Cook1980, Denley2025}. To address this challenge, we adopt a contextualized approach to effectiveness, one that is grounded in network and deterrence theory, yet also informed by the normative constraints of the CICIG–FECI alliance and the structural complexity of Guatemala’s criminal networks.

From a deterrence-theoretical perspective, arrest and prosecution are generally believed to produce a greater deterrent effect than merely increasing statutory penalties or sentence severity —particularly in contexts of limited institutional resources \cite{Mendes2004}. When severity is emphasized, effectiveness is typically assessed by whether the length of imposed sanctions reduces future criminal behavior—either by deterring the sanctioned individuals themselves (specific deterrence) or by influencing the broader population's perception of the consequences of offending (general deterrence). Beyond these core dimensions, in the context of organized crime and illicit networks, deterrence theory has evolved toward more multifaceted approaches, such as focused deterrence strategies, that combine prevention and enforcement to disrupt recruitment, cohesion, and operational capacity within criminal groups \cite{Braga2019, Denley2025}.

Understanding the long-term deterrent value of prosecutorial discretion requires examining its relationship with judicial sentencing. Although this study does not directly analyze sentencing outcomes—since such decisions are typically individualized and offer limited transparency regarding prosecutorial intent \cite{Rasmusen2009}—we acknowledge that sentencing plays a decisive role in both incapacitating offenders and shaping public perceptions of justice. This tension highlights the need to assess prosecutorial decisions not only at the level of individual cases, but also in terms of their broader systemic impact on network disruption and their potential influence on judicial outcomes.

Prosecutorial discretion—particularly in deciding whom to charge and for which offenses—can be a powerful mechanism for deterrence, especially in contexts marked by overlapping illicit enterprises \cite{coutinho2020, Tumminello2021}. In the Guatemalan case, CICIG’s mandate explicitly shaped FECI’s prosecutorial decisions, prioritizing the dismantling of illicit networks and the fight against entrenched corruption \cite{CICIG2019}. Given Guatemala’s civil law tradition, prosecutors occupy a central role in the criminal process and are expected to deter crime by shaping legal narratives and constructing compelling charges \cite{Duff2017, Colvin2019}. Their effectiveness, however, is also contingent on institutional configurations—namely, the extent to which prosecutorial offices operate independently from the judiciary or executive branch, as well as the resources at their disposal \cite{Stenning2019, McGloin2010, MerrymanandPerezPerdomo2007}. Additionally, CICIG and FECI's model assumed that effective deterrence required a coordinated institutional strategy involving not only prosecutorial innovation but also police reform, judicial appointments, and legal modernization \cite{Zamora2019, ZamudioGonzalez2021}.

In this context, we conceptualize prosecutorial discretion not merely as a procedural function but as a strategic intervention capable of reinforcing —or undermining— the deterrent capacity of the legal system. Grounded in deterrence theory, which posits the threat of punishment as a key driver of compliance \cite{Cook1980, Bielen2024}, our contribution lies in applying insights from our multilevel model to assess prosecutorial effectiveness primarily through the lens of general deterrence —specifically, by examining how prosecutorial actions shape the perceived severity of punishment among actual and potential offenders embedded within criminal networks that span public and private institutions.

Consistent with emerging criminal network research that prioritizes the disruption of functional ties and alliances sustaining organized criminal activity \cite{vonLampe2006, Albanese2018, Carrington2016, GimenezSalinas2014, Papacristos2014}, we argue that prosecutorial effectiveness should also be measured by its capacity to alter the structural dynamics of criminal networks. This perspective encompasses how prosecutors manage co-offending relationships, address multi-offender behaviors, and strategically impute offenses—particularly those involving corruption—across actors and legal frameworks.

Ultimately, we propose that a network-based approach to deterrence effectiveness requires moving beyond isolated convictions toward a systemic strategy that enables prosecutors to dismantle corrupt relationships, disrupt criminal alliances, and strategically leverage legal tools to influence network dynamics. Such an approach recognizes that prosecutors in these settings must navigate interconnected cases and complex criminal ecosystems, balancing legal discretion with institutional constraints while pursuing the strategic objectives of deterrence.

To investigate this, we introduce a socio-legal multilevel network analysis to criminal prosecution considering three interconnected network layers: (1) the criminal network that encodes how prosecutors target individuals involved in multiple co-offending relationships within and across cases; (2) the  legal network, that shows how offenses are framed by laws, distinguishing between corruption-related and other offenses, and between sever and non-sever penalties to understand prosecutorial prioritization; (3) the prosecution network that encodes how offenses are legally framed within each case to detect patterns of consistency, escalation, or strategic diversification.

\subsection{Criminal, legal and prosecution networks in Guatemala}

The case of Guatemala is particularly interesting because it represents a comparatively successful example of prosecuting complex criminal networks. The CICIG-FECI investigated and prosecuted several cases of corruption and impunity in Guatemala between 2007 and 2019. In its final report, CICIG reported having investigated more than 120 cases and identified more than 70 high-complexity criminal networks, with multiple cases interconnected through the same individuals indicted \cite{CICIG2019}.

\subsubsection{The criminal network}

Of the 120 cases investigated by CICIG-FECI, our study concentrates on eight interrelated core cases\footnote{The cases are identified by their Spanish titles: (1) ``La Línea'', (2) ``Bufete de Impunidad'', (3) ``Exdiputado Gudy Rivera'', (4) ``Cooptación del Estado'', (5) ``La Coperacha'', (6) ``Caso TCQ'', (7) ``Registro de Información Catastral: caja de pagos'', and (8) ``Caso Subordinación del poder legislativo al ejecutivo''.} for which detailed data on personal interactions were available from several reports \cite{Waxenecker2019, CICIG2019, fmm2020} and the official CICIG website\footnote{\url{https://www.cicig.org/casos-listado/}}. This sample of cases is particularly significant because it implicates high-level actors within the central government, members of Congress, and judicial officials, alongside private-sector participants such as entrepreneurs, lawyers, and accountants. Together, these cases expose the operational dynamics of criminal networks during the Patriotic Party administration and encompass a wide range of offenses, including tax fraud, corruption, money laundering, abuse of authority, illicit electoral financing, and obstruction of justice (see Appendix B for concise case descriptions).

Taken these cases together, the original database contained 296 nodes of various types (multimodal), primarily individuals, companies, and public entities \cite{Waxenecker2019}. The ties between nodes represented interactions within the illicit network, such as communications, agreements, transfers, bribes, and contracts. We transformed this data into a one-mode criminal network comprising 189 individual actors (natural persons) connected by 365 ties. Of these actors, 87 are classified as public actors, while the remaining 102 are non-public actors, primarily from the private sector. Each of these ties, regardless of their “strength” or “nature”, played a critical role in sustaining the overall criminal enterprise. By aggregating them, we capture the full spectrum of relationships that enabled information flow, resource transfers, and operational coordination within the criminal network.

\subsubsection{The legal framework network}

Out of the eight cases included, in Table \ref{tab:tab1} we identified 21 distinct offenses imputed by CICIG-FECI to criminal actors. Therefore, the legal framework described encompasses 6 laws, addressing these 21 distinct criminal offenses\footnote{Notably, “unlawful association” was charged to most of the actors across all cases. We chose to omit this offense from our dataset. First, in many criminal cases, individuals inherently interact to commit unlawful activities, making “unlawful association” less informative in revealing specific patterns of collaboration. Second, this particular offense overshadows the nuanced and meaningful relationships among other crimes and actors within our models because of its high centrality. Consequently, our network encompasses a total of 21 distinct offenses.} within the existing Guatemalan judicial system (see appendix C for a brief description of the laws). The legal framework is depicted as a one-mode network, derived from a projection of a two-mode network where laws and their corresponding offenses are interconnected. Essentially, this one-mode legal network links offenses that are governed by the same law.

\begin{table}[h!]
    \centering
    \footnotesize
    \begin{tabular}{|l|c|l|}
        \hline
        \textbf{Law} & \textbf{Count} & \textbf{Specific offenses} \\
        \hline
        \multirow{7}{*}{Penal code} & \multirow{7}{*}{7} & Illicit electoral financing \\
        & & Violation of the constitution \\
        & & Ideological falsehood \\
        & & Unregistered electoral financing \\
        & & Malfeasance (judicial misconduct) \\
        & & Extortion of public officials \\
        & & Swindle \\
        \hline
        \multirow{10}{*}{Law against corruption} & \multirow{10}{*}{10} & Passive bribery \\
        & & Active bribery \\
        & & Embezzlement \\
        & & Fraud \\
        & & Influence peddling \\
        & & Illicit enrichment \\
        & & Obstruction of criminal prosecution \\
        & & Abuse of authority \\
        & & Illegal payments \\
        & & Breach of duty \\
        \hline
        Law against money laundering & 1 & Money laundering and other assets \\
        \hline
        Law against fraud & 1 & Customs fraud \\
        \hline
        Law against organized crime & 1 & Obstruction of justice \\
        \hline
        Law against drug trafficking & 1 & Criminal association \\
        \hline
    \end{tabular}
    \caption{Offenses by law and count in the prosecution network.}
    \label{tab:tab1}
\end{table}

Additionally, according to the Guatemalan legal framework, offenses are categorized into 12 corruption-related offenses and 9 non-corruption offenses. Each offense is further categorized by severity: low severity (punishable by less than six years of imprisonment) and high severity (exceeding six years) (see Table \ref{tab:tab2}). These classifications highlight the complex landscape of illegal activities being tackled, and allows to analyze the prosecutorial strategy's depth and the legal system's capacity to address a wide spectrum of complex criminal behaviors. 

\subsubsection{The prosecution network}

The prosecution network is a two-mode network, which connects 21 legal offenses to 189 individual actors via 250 imputation ties. Table \ref{tab:tab2} shows the count of imputation ties for each of the offenses. These ties, denoted as ``imputation'', represent the prosecution process by which specific offenses are legally attributed to individual actors, reflecting their roles and activities within the criminal milieu. Each tie represents a choice by prosecutors to attribute specific criminal offenses to particular actors, reflecting their alleged roles and the extent of their involvement in the broader criminal network. Overall, of the criminal actors, 141 have been charged, whereas 48 remain uncharged. 

\begin{table}[h!]
    \centering
    \resizebox{\textwidth}{!}{
    \begin{tabular}{|l|c|l|l|c|}
        \hline
        \textbf{Offense} & \textbf{Count (imputations)} & \textbf{Offense type} & \textbf{Severity category} & \textbf{Average penalty (in years)} \\
        \hline
        Passive bribery & 53 & corruption & high & 7.5 \\
        Active bribery & 40 & corruption & high & 7.5 \\
        Money laundering and other assets & 40 & non-corruption & high & 13 \\
        Illicit electoral financing & 20 & non-corruption & high & 8 \\
        Embezzlement & 20 & corruption & high & 7.5 \\
        Customs fraud & 20 & non-corruption & high & 8.5 \\
        Fraud & 13 & corruption & high & 7.5 \\
        Influence peddling & 12 & corruption & low & 4 \\
        Illicit enrichment & 7 & corruption & high & 7.5 \\
        Violation of the constitution & 5 & non-corruption & high & 6.5 \\
        Ideological falsehood & 3 & non-corruption & low & 4 \\
        Unregistered electoral financing & 3 & non-corruption & low & 3 \\
        Obstruction of criminal prosecution & 3 & corruption & low & 4.5 \\
        Abuse of authority & 3 & corruption & low & 4.5 \\
        Malfeasance (judicial misconduct) & 2 & corruption & low & 4 \\
        Extortion of public officials & 1 & corruption & low & 4 \\
        Swindle & 1 & non-corruption & low & 2.25 \\
        Illegal payments & 1 & corruption & high & 7.5 \\
        Breach of duty & 1 & corruption & low & 4.5 \\
        Obstruction of justice & 1 & non-corruption & high & 7 \\
        Criminal association & 1 & non-corruption & high & 7 \\
        \hline
    \end{tabular}
    }
    \caption{Offenses included in the prosecution network, with count of imputations, type, severity category, and average statutory penalty.}
    \label{tab:tab2}
\end{table}

\subsection{Conceptualizing a multilevel network approach and hypothesis for analyzing prosecution practices}

We conceptualize a multilevel network, where nodes are categorized into distinct levels and with ties indicating relationships both within and between these levels. Within each level, a one-mode network is established, while between two adjacent levels, a bipartite (two-mode) network is formed to connect nodes across these levels \cite{lazega2008, wang2013, lazega2024}. 

\begin{figure}[ht!]
\centering
\includegraphics[width=0.8\textwidth]{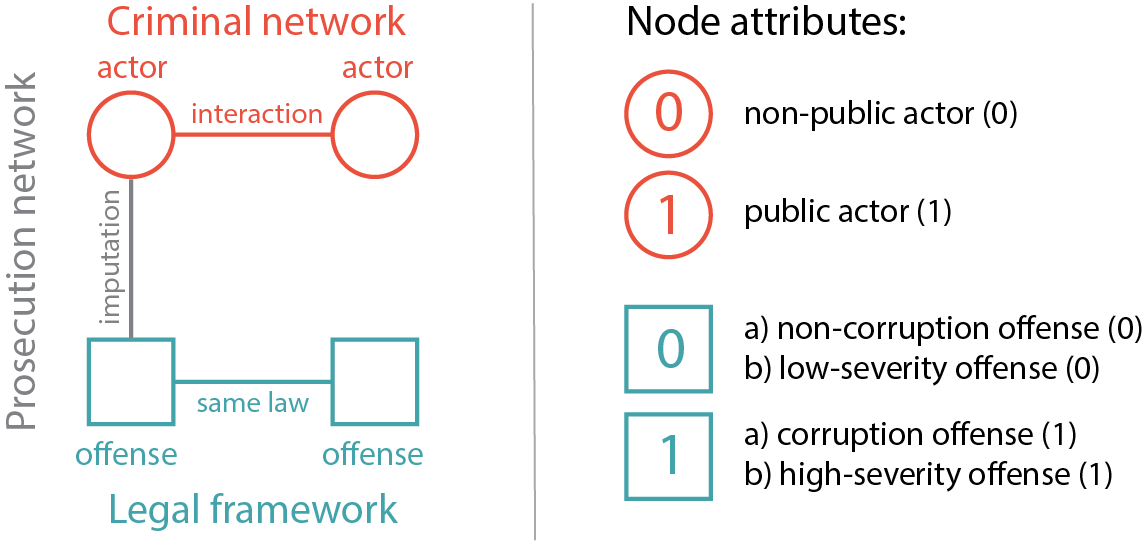}
\caption{\small Multilevel network structure consisting of: (1) a criminal network capturing within-actor ties, (2) a legal framework representing within-law connections between offenses, and (3) a prosecution network based on affiliation ties linking actors to offenses. Node attributes are encoded as follows: offense nodes are labeled by type (0 = non-corruption, 1 = corruption) and by severity (0 = low severity, 1 = high severity); actor nodes are labeled by sector (0 = non-public, 1 = public).}
\label{fig:fig2}
\end{figure}

Under the previous convention\footnote{Note that this multilevel convention is equivalent to a multilayer network in which the one-mode networks represent different layers while the bipartite networks represent the inter-layer connections.}, we combine the one-mode criminal network and the one-mode legal framework as two levels, and the bipartite prosecution network connecting nodes between these levels. Figure \ref{fig:fig2} depicts this multilevel network approach, where the circles represent individuals actively involved in the criminal network, while the squares denote criminal offenses defined by law. Ties in the criminal network and in the legal framework are called \textit{within-ties} and ties in the prosecution network are known as \textit{affiliations}.

The criminal network includes nodes that signify individual actors involved in criminal activities. In the legal framework, nodes correspond to specific offenses as delineated by the legal system. Additionally, the offense nodes have attributes indicating whether the offenses are related to corruption or not. The bipartite prosecution network acts as a critical level that links offense nodes from the legal framework to actors within the criminal network, and effectively bridges the one-mode levels of our model. In other words, the bipartite network reflects the prosecutorial decisions to impute particular offenses to specific individuals, guided by the available evidence and the overarching prosecution strategy. This multilevel approach moves away from focusing on isolated elements, like criminal actors or specific offenses, and instead emphasizes the complex interdependencies (ties) within and between various levels.

Recent research has demonstrated the effectiveness of employing a minimal building block approach for a theoretically informed empirical analysis of multilevel networks within various fields, including environmental governance networks \cite{bodin2016theorizing, barnes2017, barnes2022}. In these studies, social-ecological building blocks are identified as minimal sets of nodes (actors and ecological resources) and ties (their interdependency) that represent critical configurations, capturing essential patterns of social, ecological, and socio-ecological interdependency. Incorporating this concept into our study, we define prosecution building blocks as key configurations to understand and analyze prosecution strategies and practices to deter criminal networks. In general, this approach allows theoretical linking of the building blocks to specific processes, challenges, and hypotheses \cite{bodin2016theorizing}. 

\begin{figure}[ht!]
\centering
\includegraphics[width=0.9\textwidth]{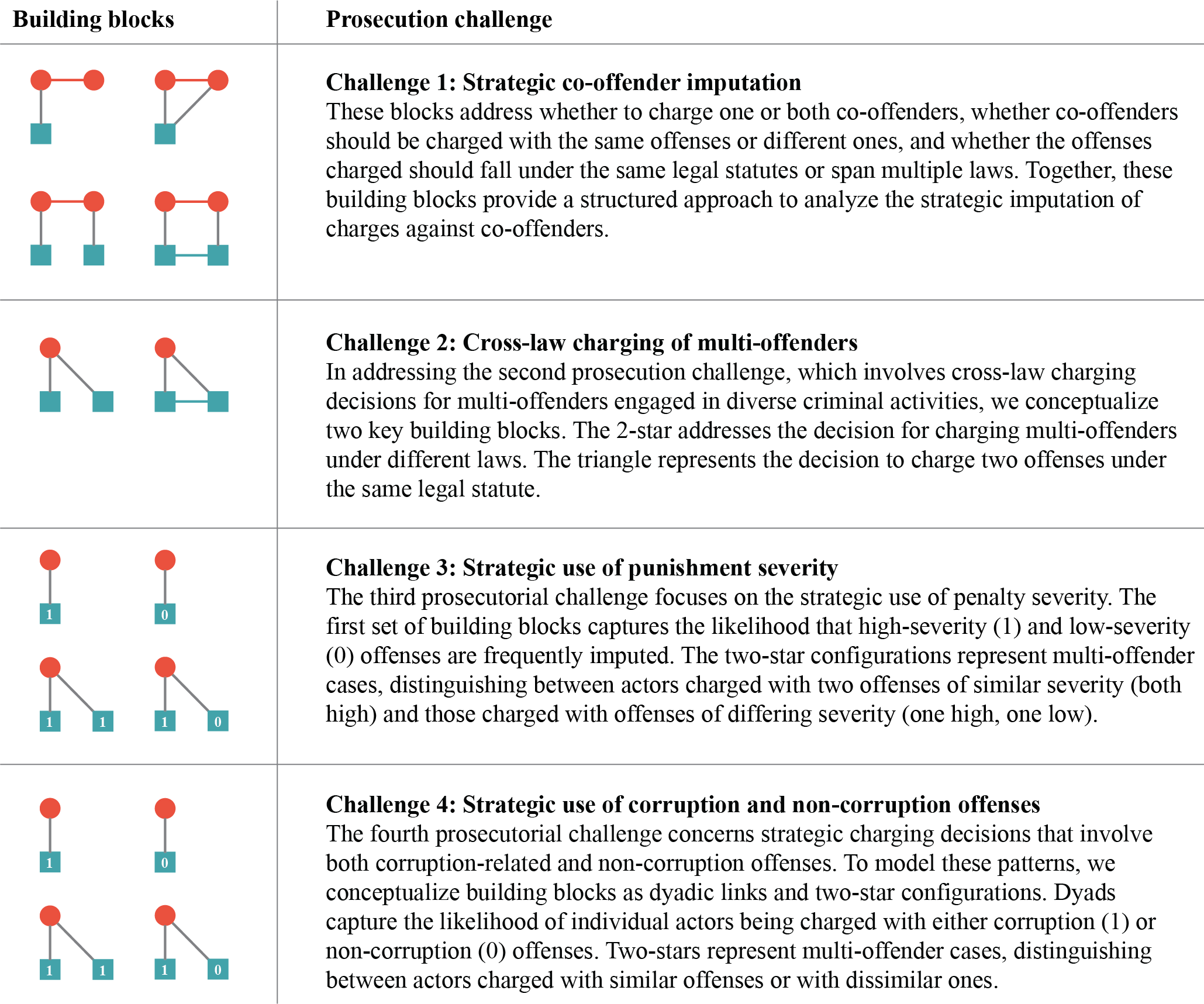}
\caption{\small Prosecution challenges and building blocks. Red circles depict criminal actors, while blue squares denote offenses.}
\label{fig:fig5}
\end{figure}

In our research, these challenges, hypotheses, and building blocks focus specifically on the prosecution of co-offenders and multi-offenders within the criminal network, analyzing significant patterns within the multilevel network. We identify four key challenges for more effective prosecution strategies against criminal networks (see Figure \ref{fig:fig5}): 

\begin{enumerate}
    \item How can co-offenders be prosecuted more effectively within complex criminal networks?
    \item How can multi-offenders be efficiently prosecuted through the strategic application of multiple legal statutes?
    \item How can prosecutorial strategies maximize deterrence by prioritizing offenses with higher statutory penalties?
    \item How can the prosecution of criminal actors be optimized by combining corruption-related and non-corruption offenses?
\end{enumerate}

These challenges form the cornerstone of our analysis, focusing on the effectiveness of prosecutorial strategies and practices in dismantling complex criminal networks. To address these challenges, and following our overarching argument that dismantling complex criminal networks needs complex prosecution strategies, we also require complex building blocks. Within our multilevel approach, complex building blocks are those in which criminal actors are charged with offenses that span multiple legal statutes and include a mix of both corruption and non-corruption related offenses in the context of their multiple interactions in the criminal network. In practical terms, this could mean that two directly connected individuals within the criminal network might face different charges; for example, one could be charged with bribery while the other is charged with money laundering, reflecting their distinct role within the network. Similarly, an individual engaged in multiple criminal activities might be charged under different statutes. For instance, an individual could face charges for tax evasion under financial laws and, separately, for trafficking illegal substances under drug enforcement laws. 

Overall, these meaningful building blocks and hypotheses are central to inferential network analysis and allow for the examination of localized mechanisms that contribute to emergent global features of the network \cite{robins2007}. The integration of theory and methodology allows us not only to hypothesize specific patterns of effective prosecution configurations but also to test these hypotheses against empirical data.

\subsubsection{Challenge 1/Hypothesis 1: Strategic co-offender imputations}

The \textbf{first challenge} is about prosecuting co-offenders and strategically allocating charges. Generally, the co-offender network perspective argues that most crimes are perpetrated by multiple individuals, and it is more common for offenders to commit crimes in collaboration with a co-offender \cite{Morselli2016}. In our study, co-offenders are defined as directly connected actors at distance 1 within the criminal network, that is, dyads of individuals linked by observed collaboration in illicit activities. In the Guatemalan criminal network under analysis, each actor participates on average in more than three dyads (calculated by the mean degree in Table \ref{tab:descriptives}), underlining the prevalence of collaborative criminal behavior and the structural relevance of addressing this challenge.

Prosecuting co-offenders entails strategically imputing charges, including the decision of whether to charge one or both actors involved. The decision extends to whether co-offenders should be charged with the same offenses or different ones, and whether these offenses should come from the same legal statutes or span multiple laws. This decision-making process, based on prosecutorial discretion \cite{kahn1962, chemerinsky2009, asp2019, Barno2021}, should acknowledge the full range of the legal framework, along with the extent and interconnectedness of criminal activities within complex networks, to positively influence the likelihood of success in criminal prosecution \cite{Autolitano2016, Rose-Ackerman2018, trejo2023}. Each charging decision carries significant implications for demonstrating the degree of collaboration and shared responsibility among co-offenders within a complex criminal network. The strategic imputation of co-offenders is essential for accurately reflecting their interactions and individual roles in the criminal enterprise. These prosecutorial choices not only shape the narrative of the case but also influence judicial outcomes by amplifying the impact of charging strategies on the severity and coherence of defendants’ final sentences \cite{Barno2021}.

In this challenge we conceptualize four building blocks (see Figure \ref{fig:fig5}), which represent deciding whether to charge one or both co-offenders, whether they should face identical or distinct offenses, and whether those offenses should be prosecuted under a single statute or across multiple laws. Since strategic and effective prosecution aims to indict all (or most) co-offenders, given the interdependency within cases and the judicial processes \cite{Waring1998, mcgloin2014}, we hypothesize that co-offenders are more likely to be jointly charged than to have only one of them charged, reflecting prosecutorial strategies designed to coherently target observed collaborations within the criminal network \textbf{(H1)}.

\subsubsection{Challenge 2/Hypothesis 2: Cross-law charging of multiple-offenders}

The \textbf{second challenge} concerns multi-offenders—individuals implicated in multiple, distinct criminal offenses connected by shared motivations or participation in organized structures. This differs from \textit{repeat offenders}, who commit the same type of offense repeatedly \cite{asp2019, audenaert2021}. In our study, multi-offending is operationalized through configurations in which one central node (i.e., a criminal actor) is linked to two other nodes representing distinct criminal offenses (see Figure \ref{fig:fig5}).

In Guatemala, prosecuting complex criminal networks required sophisticated, multi-layered charging strategies. Although the Law against corruption was the most frequently used statute (covering about 60 percent of all imputations), prosecutors had to decide whether to consolidate charges under a single legal framework or to apply multiple statutes that better reflected the nature and complexity of the conduct. Given CICIG–FECI’s conceptualization of these structures as political and economic illicit networks, dismantlement could not rely on a single law or offense type. Instead, it demanded the use of diverse legal provisions—particularly those addressing corruption and organized crime—to capture the multifaceted nature of the criminal activity \cite{Barno2021}.

Accordingly, we hypothesize that prosecutors strategically combined distinct offenses and legal frameworks to expand defendants’ legal exposure and heighten the perceived risk of prosecution. We expect multi-offenders to be more frequently charged with combinations of offenses falling under different statutes rather than within a single legal framework (\textbf{H2}). Such a pattern would indicate an intentional prosecutorial effort to reflect the complexity of criminal conduct and to target the structural versatility of Guatemala’s political–economic illicit networks.

\subsubsection{Challenge 3/Hypothesis 3: Strategic use of punishment severity}

The \textbf{third challenge} addresses the role of punishment severity in prosecutorial strategy. Although prosecutorial systems vary significantly across countries—with differences in institutional functions, degrees of independence, procedural norms, and exposure to political pressure influencing decision-making processes \cite{Stenning2019} —empirical research suggests that prosecutorial effectiveness is not primarily determined by conviction rates or the number of indictments. Instead, it is often shaped by the severity of the penalties pursued, as individuals tend to be more responsive to the perceived threat of substantial punishment than to the likelihood of prosecution alone \cite{Rasmusen2009, ChalfinandMcCrary2017}. For the purpose of this analysis, we classify punishments as \textit{high severity} when the maximum statutory penalty exceeds eight years of imprisonment, and as \textit{low severity} when it falls below this threshold.

Following this logic, we expect to observe a prosecutorial tendency within CICIG-FECI cases to prioritize the imputation of offenses associated with high statutory penalties as a strategic means of enhancing deterrence and increasing legal leverage against complex criminal actors (\textbf{H3}). This tendency should manifest both in dyadic actor–offense connections and in two-star configurations involving multiple high-severity charges (see Figure \ref{fig:fig5}), thereby reinforcing the potential for systemic deterrent effects.

\subsubsection{Challenge 4/Hypothesis 4: Strategic use of corruption and non-corruption offenses}

The Guatemalan criminal network is closely linked to state institutions and public officials. As a result, one might expect a predominance of corruption-related charges. However, empirical investigations reveal that actors embedded in such illicit political-economic networks often participate in a broader range of criminal conduct, encompassing both corruption-related and non-corruption offenses \cite{CICIG2019, Waxenecker2019, trejo2023}. These diverse criminal activities are not isolated; rather, corruption and other forms of economic crime often act as strategic complements within criminal networks and serve to connect different illicit markets \cite{kugler2005, Albanese2018, Joseph2021}. These crimes increase profitability, particularly in high-impunity contexts \cite{Spector2011, Buscaglia2013}, and facilitate operations that span the public and private sectors \cite{Bouchard2020, Sergi2019}.

In this context, the \textbf{fourth prosecutorial challenge} concerns how to charge multi-offenders: whether to focus on corruption-related offenses alone or to impute a combination of corruption and non-corruption offenses. Although the high salience of corruption suggests that it should be frequently imputed, we argue that the complexity and hybrid nature of the criminal network makes it more likely that prosecutors will adopt a mixed charging strategy. Specifically, we expect a greater incidence of multi-offenders charged with combinations of dissimilar offenses—spanning both corruption and non-corruption domains—than with offenses of a single type. This strategy would reflect efforts to accurately portray the multifaceted criminal roles of key actors and to leverage the legal system’s full range of prosecutorial tools.

To model these patterns, we analyze both dyadic actor–offense connections and two-star configurations where an actor is linked to two offenses, distinguishing specifically between corruption-related and non-corruption offenses. We hypothesize that, although corruption-related offenses are more frequently imputed overall, multi-offenders are more likely to be charged with combinations of dissimilar offenses than with offenses of a single type (\textbf{H4}). This reflects a prosecutorial strategy aimed at capturing the hybrid roles of actors within political-economic criminal networks and communicating the structural complexity of their conduct.

\section{Methods: multilevel ERGMs}

Previous research in criminology that examines co-offending, sentencing, and related factors using multilevel approaches has traditionally relied on linear regression techniques, particularly Hierarchical Linear Models (HLMs). These models are well-suited for analyzing the relationship between independent and dependent variables in nested data structures (e.g., individuals nested within jurisdictions) \cite{Waring1998, UlmerJohnson2004}. HLMs have been extensively used to assess outcomes such as incarceration rates, sentence length, recidivism, and the effects of race and gender on sentencing outcomes \cite{Ulmer2007, Wang2010, Kutateladze2014}. 

In contrast, Multilevel Exponential Random Graph Models (ERGMs) are designed to model the structure of relationships within networks \cite{lazega2008}. ERGMs are particularly effective for analyzing cross-sectional network data, providing a robust statistical framework in which network ties are treated as conditionally dependent. These models assume that ties between actors do not form randomly but are shaped by underlying social processes, reflected in specific local configurations or “building blocks” (also referred to as motifs, graph statistics, or network terms). Common motifs include edges, stars, triangles, and four-cycles, each representing distinct micro-level processes. In an ERGM, the occurrence of these motifs is quantified by graph statistics; when associated parameters are positive and statistically significant, it indicates that these motifs are more frequent in the observed network than would be expected by chance, given the model’s constraints \cite{robins2007, lusher2013, wang2013, cranmer2020}. These patterns are crucial for understanding how local interactions shape broader network structures.

Rather than focusing on individual outcomes, ERGMs estimate the probability of tie formation between nodes based on node-level attributes (e.g., type of offense or sector affiliation), structural features (e.g., homophily, transitivity, or triadic closure), and cross-level dependencies. Multilevel ERGMs extend this capability by enabling the simultaneous analysis of interconnected layers of networks, making them applicable across a variety of disciplines \cite{lazega2008, zappa2015, zhu2016, koskinen2023}, including the study of criminal networks \cite{coutinho2020}.

We employ Multilevel ERGMs to analyze the strategic or preferred configurations of co-offender prosecution, particularly in contexts involving both public and private actors, overlapping charges, and a combination of corruption-related and non-corruption offenses. This approach enables us to explore how prosecutorial strategies can be systematically modeled and evaluated for their potential to disrupt criminal networks.

Our analytical framework is further enriched by the conceptualization of building blocks derived from the study of complex multilevel socio-ecological systems \cite{bodin2016theorizing, barnes2017, barnes2022}. Drawing on these methods allows us to examine the nuanced interdependencies within and across levels in our network, offering a comprehensive and empirically grounded perspective on the systemic dynamics shaping prosecutorial decision-making and network resilience.

\begin{figure}[ht!]
\centering
\includegraphics[width=1\textwidth]{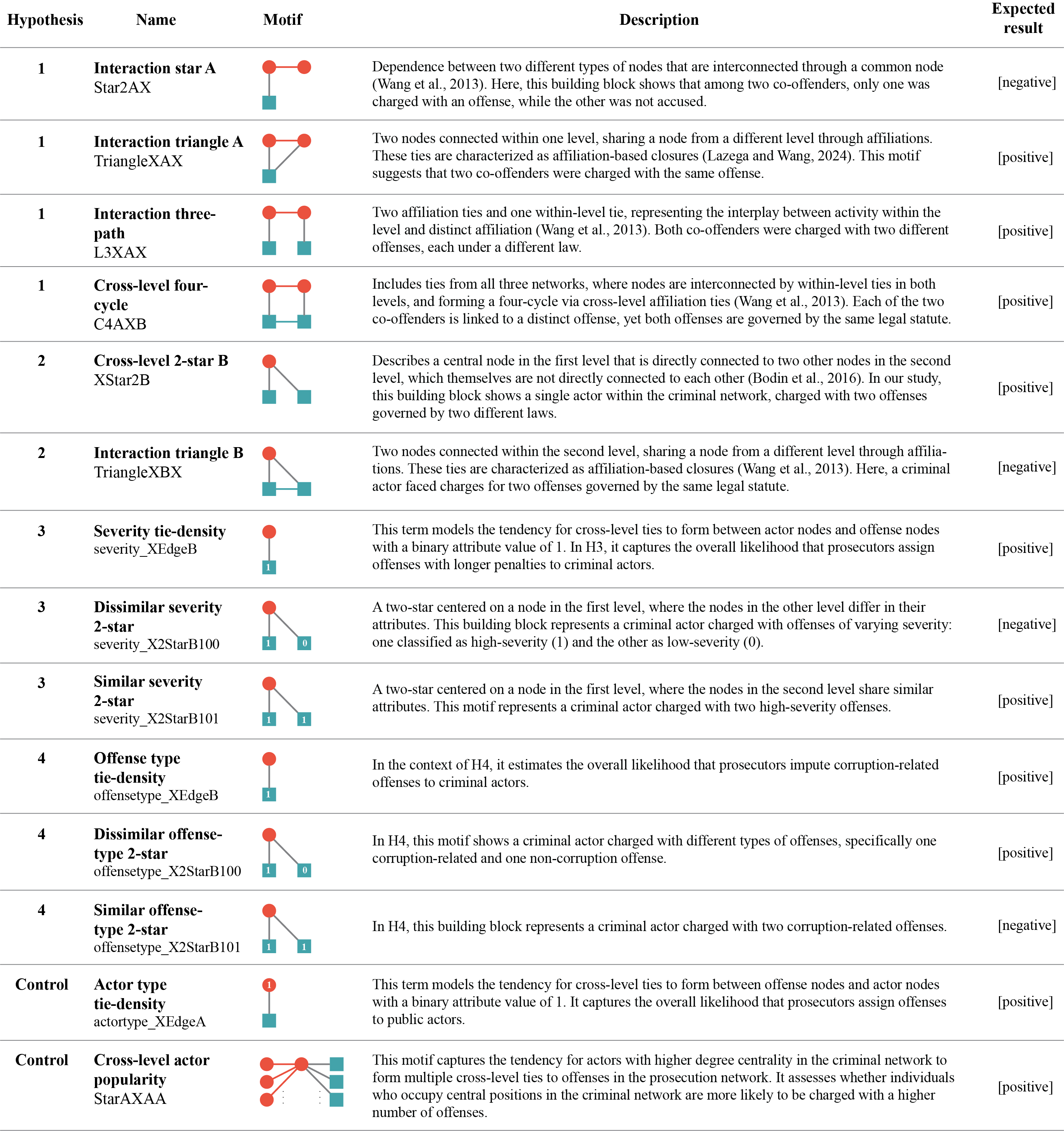}
\caption{\small Prosecution building blocks. This figure illustrates the key network motifs used in the analysis. Red circles represent criminal actors, blue squares denote offenses. Each building block is linked to a specific hypothesis and expected estimation result.}
\label{fig:fig3}
\end{figure}

Our analysis focuses on the set of 14 prosecution building blocks outlined in Figure \ref{fig:fig3}. This set of minimal prosecution building blocks capture an assumed relationship within and between the legal framework and the criminal network:

\begin{itemize}

    \item For challenge 1, the Interaction star (Star2AX) shows that among two co-offenders, only one was charged with an offense, while the other was not accused. The Interaction triangle A (TriangleXAX) suggests that two co-offenders were charged with the same offense. The Interaction three-path (L3XAX) signifies that both co-offenders were charged with two different offenses, each under a different law. Lastly, the Cross-level four-cycle or Cross-level four-cycle (C4AXB) \cite{lazega2024} signifies that each of the two co-offenders was charged with a distinct offense, yet both offenses are governed by the same legal statute. This structural differentiation within the network models provides a nuanced understanding of how charges are distributed among co-offenders in relation to the legal framework.

    \item For challenge 2, the Cross-level 2-star B (XStar2B) indicates centralization \cite{bodin2016theorizing}, where a single actor within the criminal network was charged with two offenses governed by two different laws.\footnote{Such 2-stars may also be components of higher-order configurations involving actors charged with more than two offenses.} The Interaction triangle B (TriangleXBX) configuration reveals that a criminal actor was charged with two offenses that fall under the same legal statute. This delineation helps to illustrate the range of legal actions applied to individual actors within the network.

    \item For challenge 3, the Severity tie-density term (severity-XEdgeB) captures the overall tendency to impute high-severity offenses within the prosecution network. The Similar severity 2-star (severity-X2StarB101) indicates cases where a criminal actor was charged with two high-severity offenses, while the Dissimilar severity 2-star (severity-X2StarB100) reflects combinations offenses with differing levels of severity. Together, these building blocks assess the strategic use of punishment severity as a means to reinforce potential deterrent effects within complex criminal networks.

    \item For challenge 4, the Offense type tie-density term (offensetype-XEdgeB) captures the overall tendency to impute corruption offenses. The Similar offense type 2-star (offensetype-X2StarB101) indicates that a criminal actor was charged with the same type of offense. Conversely, the Dissimilar offense type 2-star (offensetype-X2StarB100) shows that a criminal actor was charged with different types of offenses, specifically one corruption-related offense and one non-corruption offense. This distinction highlights the diversity in the prosecutorial approach to individual actors based on the nature of their alleged criminal activities.

\end{itemize}

Our model includes two control terms related to actor characteristics. The Actor type tie-density term (actortype-XEdgeA) captures the baseline tendency for imputations based on whether the node represents a public or non-public actor. The Cross-level actor popularity term (StarAXAA) assesses whether individuals who occupy central positions in the criminal network are more likely to be charged with a higher number of offenses in the prosecution network, thereby reflecting the concentration of prosecutorial focus on more prominent actors.

And finally, to model endogenous network effects that could explain the presence of a tie, we include network terms to control for affiliation-tie density (XEdge), affiliation-popularity of offenses (XASB), triadic multilevel closure (ATXBX), and Cross-level 2-star A (XStar2A) \cite{wang2013, koskinen2023}. 

The computations were done using the \textit{Program for the Simulation and Estimation of (p*) Exponential Random Graph Models for Multilevel Networks} (MPNet) \cite{wang2022mpnet}, and as such, the terminology for naming network motifs or building blocks (e.g., L3XAX or Star2AX) is adopted from this software\footnote{In MPNet, the first one-mode network representing the criminal network, the second one-mode network representing the legal framework, and the bipartite network representing the prosecution network.}. We implemented a step-wise estimation strategy: starting with a baseline model including the endogenous and control terms (Model 0), and progressively introducing additional terms corresponding to each hypothesis in subsequent models (Model 1 to Model 4). We also conducted the goodness-of-fit (GOF) analysis to assess the model's accuracy and applicability.

\section{Results}

\subsection{Descriptive results}

Table \ref{tab:descriptives} presents the descriptive properties of the different network levels. The criminal network (A) consists of 189 actors connected by 365 ties, with a very low density of 0.021 and a mean degree of 3.86, indicating that ties are sparse and unevenly distributed, as reflected in the relatively high degree variation (SD = 7.13). By contrast, the legal framework (B) of 21 offenses is substantially denser, with a density of 0.314 and an average degree of 6.29.

\begin{table}[htbp]
\centering
\footnotesize
\begin{tabular}{llccccc}
\toprule
\textbf{Level} & & \textbf{n\_nodes} & \textbf{n\_edges} & \textbf{Density} & \textbf{Mean degree} & \textbf{SD degree} \\
\midrule
A: Criminal & & 189 & 365 & 0.021 & 3.86 & 7.13 \\
B: Legal framework & & 21 & 66 & 0.314 & 6.29 & 3.41 \\
\multirow{2}{*}{X: Prosecution network} 
  & Actors     & 189 & \multirow{2}{*}{250} & \multirow{2}{*}{0.063} & 1.32  & 1.25 \\
  & Offenses   & 21  &                      &                        & 11.91 & 15.27 \\
\bottomrule
\end{tabular}
\caption{Descriptive statistics for the criminal (A), legal framework (B), and prosecution (X) networks. For the prosecution network (X), degree statistics are reported separately for actors and offenses.}
\label{tab:descriptives}
\end{table}

The prosecution network (X), linking actors to offenses, contains 250 ties and has a density of 0.063. Within this bipartite structure, degree distributions differ strongly between node types: actors are on average connected to only 1.3 offenses, whereas offenses are linked to nearly 12 actors on average. Of all actors in the network, 32.3\% are identified as multi-offenders, 42.3\% are linked to a single offense, while the remaining 25.4\% are not imputed at all. Table \ref{tab:laws} summarizes the distribution of prosecutorial ties, disaggregated by law, offense type and severity category. The Law against Corruption dominates the prosecutorial framework, encompassing 10 different offenses and accounting for 153 imputations (over 60 percent of the total). Strikingly, all of these imputations fall under corruption offenses, with the vast majority classified as high-severity (134 out of 153). Among non-corruption statutes, the Law against Money Laundering (40 imputations) and the Law against Fraud (20 imputations) stand out, with imputations concentrated entirely in the high-severity category. The Penal Code also contributes substantially (35 imputations), but unlike the specialized statutes, it combines both non-corruption (32) and corruption offenses (3), with a more balanced distribution across severity categories. In contrast, the Law against Organized Crime and the Law against Drug Trafficking are rarely invoked, each associated with only a single imputation.

\begin{table}[h!]
\centering
\resizebox{\textwidth}{!}{%
\begin{tabular}{lcccccc}
\hline
\textbf{Laws} & \textbf{No. of offenses} & \textbf{Total X-ties} & \multicolumn{2}{c}{\textbf{Offense type}} & \multicolumn{2}{c}{\textbf{Severity category}} \\
 & & & Non-corruption & Corruption & Low & High \\
\hline
Law against corruption & 10 & 153 & 0 & 153 & 19 & 134 \\
Law against money laundering & 1 & 40 & 40 & 0 & 0 & 40 \\
Penal code & 7 & 35 & 32 & 3 & 10 & 25 \\
Law against fraud & 1 & 20 & 20 & 0 & 0 & 20 \\
Law against drug trafficking & 1 & 1 & 1 & 0 & 0 & 1 \\
Law against organized crime & 1 & 1 & 1 & 0 & 0 & 1 \\
\hline
\textbf{Total general} & \textbf{21} & \textbf{250} & \textbf{94} & \textbf{156} & \textbf{29} & \textbf{221} \\
\hline
\end{tabular}}
\caption{Distribution of prosecution ties by law: number of imputations, and classifications by offense type and severity category.}
\label{tab:laws}
\end{table}

\begin{figure}[ht!]
\centering
\includegraphics[width=1\textwidth]{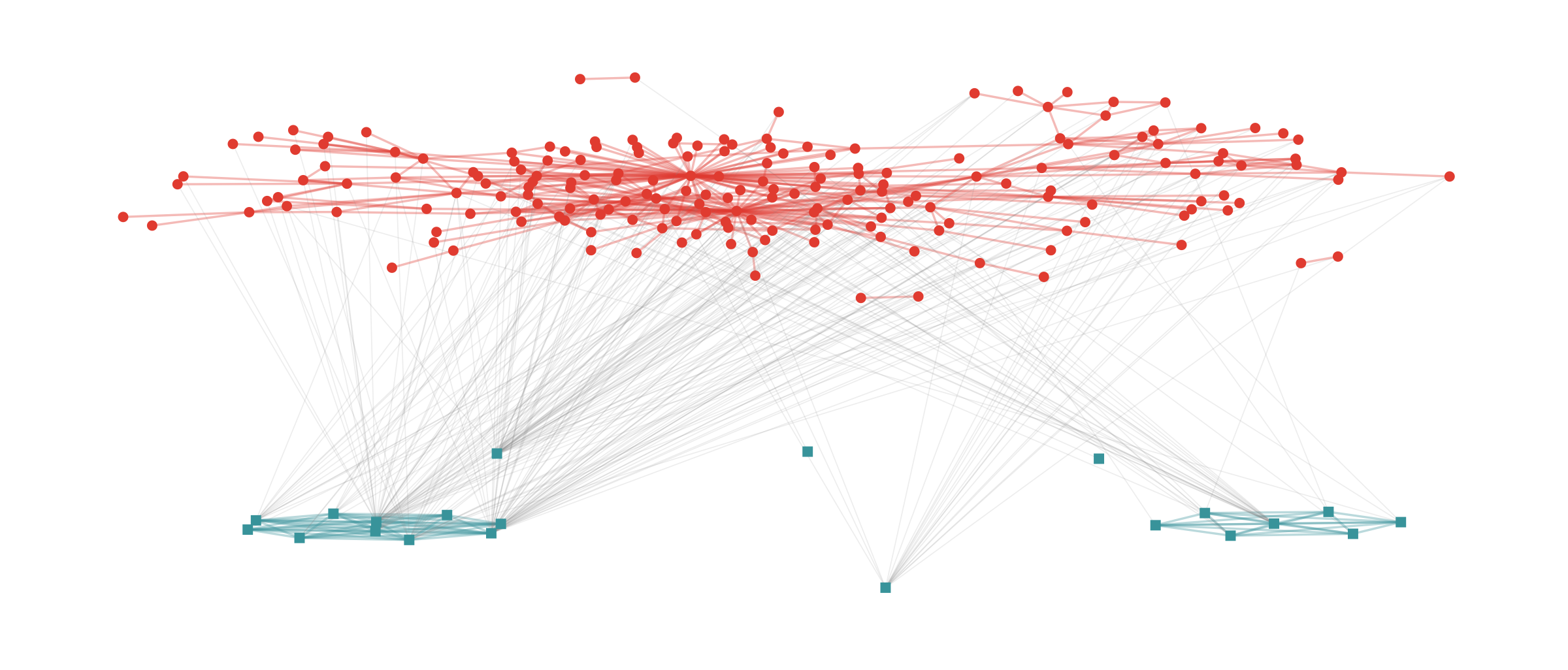}
\caption{\small Multilevel network. This figure depicts the integrated multilevel network that combines three distinct but interrelated layers. The upper layer represents the criminal network (red circles and red ties), the lower layer captures the legal framework (blue squares and blue ties), and the cross-level layer, shown with grey ties, constitutes the prosecution network.}
\label{fig:fig4}
\end{figure}

Finally, as depicted in Figure \ref{fig:fig4}, the integration of all three levels produces a cohesive structure in which prosecutorial ties bridge the criminal network with the legal framework, creating the backbone of the multilevel network.

\subsection{Results from the inferential multilevel analysis}

We examine the existing criminal network represented as a one-mode actor-to-actor network, alongside the established legal framework depicted as a one-mode offense-to-offense network. Prosecutors play a pivotal role by linking offenses to criminal actors, creating a two-mode prosecution network. Our primary interest lies in this two-mode network to discern the underlying prosecution strategies based on our specific building blocks. Consequently, within our model, we treat the legal framework and the criminal network as constants, focusing our analytical efforts on the two-mode prosecution network. Thus, while the one-mode layers are held constant, the two-mode layer is actively modeled when estimating the ERGM.

The multilevel ERGM results presented in Table \ref{tab:tab3} indicate that some building blocks exhibit greater statistical significance compared to others across all models. This suggests that specific network configurations are more influential in shaping the structure of the observed multilevel network.

\begin{table}[htbp]
\centering
\begin{adjustbox}{width=\textwidth}
\footnotesize
\begin{tabular}{llcccccccccc}
\toprule
\textbf{Term} & & \multicolumn{2}{c}{\textbf{Model 0}} & \multicolumn{2}{c}{\textbf{Model 1}} & \multicolumn{2}{c}{\textbf{Model 2}} & \multicolumn{2}{c}{\textbf{Model 3}} & \multicolumn{2}{c}{\textbf{Model 4}} \\
 & & Param & SD & Param & SD & Param & SD & Param & SD & Param & SD \\
\midrule
\multicolumn{12}{l}{\textit{Endogenous}} \\
Affiliation tie-density & & -11.660* & 0.732 & -8.447* & 0.722 & -8.304* & 0.687 & -8.147* & 0.862 & -8.228* & 0.830 \\
Affiliation popularity of offenses & & 3.268* & 0.285 & 2.051* & 0.256 & 1.695* & 0.240 & 1.208* & 0.344 & 1.204* & 0.352 \\
Triadic multilevel closure & & -0.392 & 0.231 & -0.124 & 0.213 & 0.008 & 0.257 & 0.175 & 0.295 & 0.241 & 0.359 \\
Cross-level 2-star A & & -0.018 & 0.101 & -0.032 & 0.110 & -0.056 & 0.105 & 0.048 & 0.581 & -0.224 & 0.616 \\
\midrule
\multicolumn{12}{l}{\textit{Control}} \\
Actor type tie-density & & 0.448* & 0.140 & 0.429* & 0.145 & 0.463* & 0.144 & 0.455* & 0.142 & 0.451* & 0.142 \\
Cross-level actor popularity & & 1.434* & 0.297 & 0.828* & 0.342 & 0.926* & 0.312 & 0.945* & 0.352 & 0.949* & 0.322 \\
\midrule
\multicolumn{12}{l}{\textit{Hypothesis 1}} \\
Interaction star A & &  &  & -0.541 & 0.033 & -0.032 & 0.023 & -0.035 & 0.021 & -0.037 & 0.024 \\
Interaction triangle A & &  &  & 0.818* & 0.055 & 0.572* & 0.061 & 0.524* & 0.065 & 0.512* & 0.065 \\
Interaction three-path & &  &  & -0.014 & 0.020 & -0.003 & 0.016 & 0.001 & 0.014 & 0.006 & 0.016 \\
Cross-level four-cycle & &  &  & 0.004 & 0.009 & 0.005 & 0.020 & 0.007 & 0.018 & -0.002 & 0.022 \\
\midrule
\multicolumn{12}{l}{\textit{Hypothesis 2}} \\
Cross-level 2-star B & &  &  &  &  & 0.022* & 0.002 & 0.021* & 0.002 & 0.021* & 0.002 \\
Interaction triangle B & &  &  &  &  & -0.036 & 0.183 & -0.069 & 0.014 & -0.107 & 0.328 \\
\midrule
\multicolumn{12}{l}{\textit{Hypothesis 3}} \\
Severity tie-density & &  &  &  &  &  &  & 0.955* & 0.335 & 0.988* & 0.338 \\
Dissimilar severity 2-star & &  &  &  &  &  &  & 0.1092 & 0.6458 & 0.1446 & 0.6433 \\
Similar severity 2-star & &  &  &  &  &  &  & -0.3139 & 0.1967 & -0.3155 & 0.1917 \\
\midrule
\multicolumn{12}{l}{\textit{Hypothesis 4}} \\
Offense type tie-density & &  &  &  &  &  &  &  &  & 0.098 & 0.176 \\
Dissimilar offense type 2-star & &  &  &  &  &  &  &  &  & 0.358 & 0.317 \\
Similar offense type 2-star & &  &  &  &  &  &  &  &  & -0.196 & 0.292 \\
\bottomrule
\end{tabular}
\end{adjustbox}
\caption{Results from the multilevel ERGM: parameter estimates and standard errors. Results that are statistically significant are denoted with an asterisk (*).}
\label{tab:tab3}
\end{table}

Testing for H1 related to strategic co-offender imputation, the model presents mixed results across the four building blocks. The Interaction star (Star2AX) does not show a significant influence on the network structure. Conversely, the Interaction triangle A (TriangleXAX) displays a significant positive estimate, underscoring the importance of triangular configurations involving two co-offenders and one offense. This suggests that prosecutors tend to impute the same offense to directly connected co-offenders.

However, when examining more complex configurations such as the Interaction three-path (L3XAX) and the Cross-level four-cycle (C4AXB), the findings reveal very small and statistically insignificant parameter estimates. This indicates that these building blocks do not have a significant impact on the network’s structure, suggesting that simpler configurations might play a more critical role in prosecuting co-offenders.

Testing for H2, the results align with our expectations. The Cross-level 2-star B (XStar2B) shows positive and significant estimates, indicating that individuals tend to be linked to multiple offenses from different laws. Additionally, the Interaction triangle B (TriangleXBX) yields a negative estimate, suggesting a minimal and non-significant tendency for building blocks where one criminal actor is linked to two offenses under the same law. 

Testing for H3, the results provide partial support. The severity tie-density term (severity\_XEdgeB) is positive and statistically significant, indicating a clear prosecutorial preference for imputing high-severity offenses. In contrast, the dissimilar severity two-star term (severity\_X2StarB100) is positive but not statistically significant, providing no strong evidence that prosecutors systematically combine high- and low-severity offenses when charging multi-offenders. Likewise, the similar severity two-star term (severity\_X2StarB101) is negative and not significant, suggesting that multi-offenders are not more likely to be charged with two high-severity offenses.

Testing for H4, the findings fail to support the hypothesis. While the offense type tie-density term (offensetype\_XEdgeB) is positive, indicating a tendency to impute corruption-related offenses, and the dissimilar offense type two-star term (offensetype\_X2StarB100) is also positive, suggesting a potential inclination to combine corruption and non-corruption charges for multi-offenders, neither reaches statistical significance. Similarly, the similar offense type two-star term (offensetype\_X2StarB101) is negative and non-significant, indicating no systematic tendency to charge multi-offenders with two offenses of the same type.

The control terms, actor type tie-density (actortype-XEdgeA) and cross-level actor popularity (StarAXAA), produce consistently positive and statistically significant estimates across all models.

The analysis of endogenous configurations reveals several consistent findings in the ERGM results. The parameter for affiliation-tie density (XEdge) shows a statistically significant negative effect, indicating that, other factors being constant, the formation of additional prosecution ties within the network is generally unlikely. In contrast, the parameter for affiliation-popularity of offenses (XASB) demonstrates a significant positive effect, suggesting a preference for more frequently implicated offenses in the prosecution network. Meanwhile, neither the triadic multilevel closure term (ATXBX) nor the two-star motif (XStar2A) is statistically significant, indicating limited evidence of structural patterns involving clustering or lower-order configurations centered on offenses.

Finally, all models have successfully converged and the overall fit of Model 4, measured by the Mahalanobis distance of 693, is satisfactory \cite{lusher2013}. Additionally, the graph statistics incorporated into the model along with the global configurations that represent the network structure (such as standard deviation and skewness of the degree distributions and the global clustering coefficient) are all well fitted (see Table \ref{tab:tab4}). A good fit is achieved when the t-ratios for included graph statistics are less than 0.1 in absolute value, and for global statistics, they are below 2.0 \cite{wang2022mpnet}.

\begin{table}[htbp]
\centering
\footnotesize
\begin{tabular}{lrrrr}
\toprule
\textbf{Statistics} & \textbf{Observed} & \textbf{Mean} & \textbf{StdDev} & \textbf{t-ratio} \\
\midrule
XEdge & 250.00 & 262.46 & 57.47 & -0.2167 \\
XStar2A & 187.00 & 204.72 & 88.74 & -0.1997 \\
XStar2B & 3696.00 & 4351.56 & 2989.01 & -0.2193 \\
XASB & 431.16 & 455.59 & 111.25 & -0.2196 \\
actortype\_XEdgeA & 144.00 & 150.37 & 31.61 & -0.2015 \\
offensetype\_XEdgeB & 156.00 & 162.66 & 40.41 & -0.1649 \\
offensetype\_X2StarB100 & 167.00 & 181.98 & 78.39 & -0.1911 \\
offensetype\_X2StarB101 & 62.00 & 66.60 & 30.38 & -0.1514 \\
severity\_XEdgeB & 221.00 & 232.01 & 48.56 & -0.2267 \\
severity\_X2StarB100 & 184.00 & 201.47 & 86.15 & -0.2028 \\
severity\_X2StarB101 & 135.00 & 147.13 & 57.84 & -0.2096 \\
Star2AX & 1538.00 & 1607.27 & 371.46 & -0.1865 \\
StarAXAA & 1317.66 & 1339.59 & 102.71 & -0.2136 \\
TriangleXAX & 223.00 & 246.14 & 107.13 & -0.2160 \\
L3XAX & 1617.00 & 1778.10 & 813.52 & -0.1980 \\
TriangleXBX & 59.00 & 64.08 & 27.87 & -0.1823 \\
ATXBX & 32.21 & 33.81 & 9.81 & -0.1629 \\
C4AXB & 457.00 & 489.28 & 228.22 & -0.1414 \\
stddev\_degreeA & 7.39 & 7.39 & -- & -1.0000 \\
skew\_degreeA & 7.67 & 7.67 & -- & -1.0000 \\
clusteringA & 0.08 & 0.08 & -- & -1.0000 \\
stddev\_degreeX\_A & 1.25 & 1.23 & 0.15 & 0.1083 \\
skew\_degreeX\_A & 1.32 & 0.91 & 0.24 & 1.6619 \\
stddev\_degreeX\_B & 15.27 & 15.39 & 6.30 & -0.0179 \\
skew\_degreeX\_B & 1.38 & 1.85 & 0.87 & -0.5373 \\
clusteringX & 0.22 & 0.15 & 0.04 & 1.5947 \\
stddev\_degreeB & 4.69 & 4.69 & -- & -1.0000 \\
skew\_degreeB & 0.99 & 0.99 & -- & -1.0000 \\
clusteringB & 1.00 & 1.00 & -- & NaN \\
\bottomrule
\end{tabular}
\caption{Test results for goodness of fit (GOF). The first column lists the configurations used in the GOF simulation; the second column shows their counts in the observed network; the third column presents the means of the simulated graph statistics; the fourth column details the standard deviations, and the fifth column displays the t-ratios \cite{wang2022mpnet}. All of these values indicate a good fit of our multilevel model.}
\label{tab:tab4}
\end{table}

\section{Discussion}

The network configurations underpinning prosecutorial imputation strategies have received limited theoretical and empirical attention \cite{Albonetti1987, Rasmusen2009}. This paper advances methodological and theoretical contributions by applying an inferential, network-based approach to the study of criminal networks. Grounded in the intersection of deterrence theory and network science, our research examines the prosecutorial action in deterring complex criminal structures. The conceptual framework integrates organized crime deterrence theory, prosecutorial discretion, and effectiveness with insights from the literature on network analysis and organized crime.

Building on insights into multilevel network analysis in socio-ecological systems \cite{bodin2016theorizing, barnes2017, barnes2022, lazega2008}, we extend these tools to a novel empirical domain: prosecutorial strategies in complex corruption cases. Specifically, we conceptualize and employ network motifs to capture relational patterns across interconnected layers—criminal actors, legal configurations that structure prosecutorial decision-making. By embedding prosecutorial actions within a multilevel network architecture, our approach provides a framework for analyzing prosecution as a systemic intervention aimed at disrupting and weakening the organizational dynamics of illicit political–economic networks. In doing so, we position network inference as a bridge between the legal-institutional study of prosecution and the relational analysis of organized crime.

Our analysis identifies recurring patterns in prosecutorial decision-making that reveal strategic adaptations to a highly criminalized environment. These adaptations, particularly efforts to dismantle corruption-enabled criminal networks, reflect deliberate choices that extend beyond the logic of individual cases, aligning with a broader deterrence-oriented prosecutorial strategy. The multi-level model sheds light on the strategic dimension of deterrence in prosecutorial practice, telling a mixed yet coherent story across the different hypotheses.

Overall, one of the most significant findings from our models concerns multi-offenders, individuals implicated in multiple and distinct criminal offenses. Our findings indicate that multi-offenders were more often linked to offenses falling under different legal statutes rather than confined to a single law. The reliance on multiple legal frameworks reflects a prosecutorial strategy designed not only to target individual acts of corruption, but also to expose and weaken the systemic connections that allowed such networks to thrive. This pattern aligns with prior research \cite{Barno2021} and reflects a deliberate prosecutorial strategy that emerged during CICIG’s early years, when initial efforts focused on advancing reforms to remove political and procedural barriers to its work. Beginning in late 2013, five priority criminal phenomena were defined: smuggling, administrative corruption, judicial corruption, drug trafficking and money laundering, and illicit electoral financing \cite{CICIG2019}. Consistent with these priorities, our study finds that most charges were brought under the Law against Corruption (153), followed by the Law against Money Laundering (40), the Penal Code (35), the Law against Fraud (20), and other statutes (2) (see Table \ref{tab:laws}). These prosecutorial lines of investigation targeted entrenched forms of organized crime, emphasizing the links between illicit funds, corruption, and political influence. Importantly, addressing multi-offenders under multiple legal frameworks proved effective not only in capturing the multifaceted nature of criminal behavior but also in ensuring broader and more comprehensive legal coverage in the fight against corruption \cite{Albanese2018, Joseph2021}.

From a deterrence perspective, combining distinct offenses under different laws increased defendants’ legal exposure and heightened the perceived risks of prosecution. In this sense, the prosecution of multi-offenders illustrates an adaptive use of the legal framework: rather than relying on narrow charges, prosecutors pursued diverse imputation strategies to reflect the breadth of the networks’ illicit practices and to expand the scope of accountability.

This effect is particularly pronounced for actors occupying central positions within the criminal network. Our results show that individuals with higher centrality in the criminal network tend to attract greater prosecutorial attention. This finding is consistent with criminal network studies \cite{Morselli2010, Diviak2018} and reflects the tendency of prosecutors, in Guatemala and elsewhere, to concentrate limited resources on cases with strong evidentiary support \cite{Rasmusen2009} and on high-profile actors.

Our model further controls and underscores the central role of public actors within prosecutorial strategies, showing that prosecution ties are more likely directed toward state officials rather than private actors. This pattern is consistent with CICIG’s conceptualization of Guatemala’s \textit{redes político-económicas ilícitas} (RPEI). These illicit political–economic networks fundamentally depend on the involvement of state actors, whose institutional positions provide the authority and access necessary to extract, channel, and redistribute resources through corrupt exchanges. In the Guatemalan context, this logic translated into a prosecutorial focus on high-ranking officials during the \textit{Partido Patriota} administration (2012–2015), whose dense relational embeddedness made them both critical enablers of illicit exchanges and highly exposed to legal action. The collapse of this administration in 2015, culminating in the resignation of the president, vice-president, and cabinet members, illustrates how prosecutorial interventions against central actors can disrupt network stability, but also how such interventions occur only under extraordinary political circumstances.

Nevertheless, the effectiveness of prosecution rests not only on the number of charges brought forward but also on the severity of the penalties attached to them. CICIG’s preference for imputing high-severity offenses was consistent with its broader mandate to combat impunity and its pedagogical objective of advancing legal doctrine and jurisprudence on diverse forms of criminality, while simultaneously strengthening institutional technical capacities, as emphasized in its final report \cite{CICIG2019}. Against this background, CICIG and FECI’s deterrence strategy weighed both the probability of uncovering links to illicit networks and the likelihood of litigation success, deliberately using high-severity charges to maximize deterrent impact and shield cases from procedural challenges. However, our results also reveal a nuanced pattern. While prosecutors clearly prioritized high-severity offenses, we find no statistical evidence that multi-offenders were deliberately charged with two such offenses. This runs counter to our theoretical expectation and suggests that, although severity was strategically employed to enhance deterrence, prosecutors did not consistently extend this strategy to the design of multi-offense charges.

Our theoretical expectation also posited that multi-offenders embedded in Guatemala’s illicit political–economic networks would be systematically prosecuted through hybrid charging strategies, combining corruption and non-corruption offenses. The empirical results, however, provide no support for this hypothesis. Prosecutors did not exhibit a consistent pattern of employing offense-type combinations, suggesting the absence of a systematic hybrid strategy in addressing multi-offenders.

Considered together, these findings indicate that CICIG–FECI’s prosecutorial approach combined a strong reliance on deterrence through the prioritization of high-severity offenses with a more selective and ultimately unsystematic use of offense-type combinations. This dual prosecutorial pattern illustrates both the opportunities and constraints in tackling entrenched political–economic networks. High-penalty charges represent a relatively straightforward means of enhancing deterrence and reinforcing prosecutorial leverage. By contrast, the integration of corruption and non-corruption offenses emerges as a fragile and context-dependent practice, likely conditioned by legal, institutional, and political limitations on how far prosecutors could extend their charging strategies.

In addition, prosecutorial strategies toward co-offenders yield a mixed set of results. Among the four modeled building blocks, only the interaction triangle term, in which two co-offenders are jointly linked to a common offense, shows a significant positive effect. The implications are twofold. First, the positive estimate for triangular configurations underscores the importance of joint imputation as a prosecutorial tool. This reflects a pragmatic strategy: leveraging shared evidence and procedural efficiencies in addressing co-offenders, particularly as FECI, burdened with a high volume of cases and defendants, sought to secure outcomes in the shortest possible time. Second, the absence of significant effects for the other building blocks indicates that prosecutors did not systematically extend this strategy to other complex configurations. The observed network exhibits lower density of prosecution ties than a random network, which further constrains the range of “building blocks” available for more sophisticated imputation strategies. Density would increase if “unlawful association” were included (see footnote 3), but meaningful building blocks require a diverse mix of charges rather than on a single, overly dominant offense. From a deterrence-network perspective, combining different building blocks, such as TriangleXAX, L3XAX, and C4AXB, could theoretically generate a “robust structure” of imputation, where co-offenders and their offenses form interconnected patterns spanning multiple cases, statutes, and charges. Such structures not only capture the complexity of the criminal network but also enhance prosecutorial capacity to disrupt corrupt alliances and frame their operations as systematic and organized \cite{trejo2023, Autolitano2016}. However, in practice, applying this strategy consistently was also limited by the heterogeneous positions of co-offenders within the network. Not all were equally positioned to engage in corruption-related crimes such as bribery, embezzlement, or fraud, alongside non-corruption charges like money laundering or illicit electoral financing.
 
\subsection{Practical implications}

Our study underlines several practical implications for prosecutorial strategies in dismantling entrenched political–economic criminal networks.

First, on actor selection, prosecuting central actors can be particularly disruptive in contexts where illicit networks are deeply embedded within and beyond the state. Because these individuals occupy brokerage positions that sustain flows of illicit exchange, prioritizing them provides great leverage for destabilizing network structures. In the Guatemalan case, this dynamic was especially visible among public officials, whose institutional authority made them pivotal nodes in political–economic illicit networks \cite{CICIG2019, Waxenecker2019}. Yet this should not be confined to public actors alone: if prosecutions focus exclusively on state officials, the broader structure may remain intact, as new officials can readily move in to occupy their vacated roles. Effective and persistent disruption therefore requires a strategy that addresses both public and private brokers to weaken the systemic foundations of illicit networks.

Second, on the number and severity of offenses, deterrence is not achieved by quantity alone but by strategic choices. Prosecutors leaned heavily on high-severity offenses, which increase legal leverage, but they balanced this against the risks of overcomplication when combining multiple charges. We propose that severity becomes effective when it disrupts the functional ties that support criminal activity, both through long-term punishment and by increasing the perceived threat of sanctions among co-offenders.

Third, on offense type selection, corruption-related charges dominate but leave parts of the networks’ operations under-addressed. We propose that disrupting entrenched criminal networks requires targeting both categories of offenses. Prosecuting corruption and non-corruption conduct in tandem not only disrupts immediate criminal activities—particularly when directed at public figures and highly central actors—but also undermines the structural and financial mechanisms sustaining impunity, since co-offenders within a case, regardless of their role, benefit from the broader environment of protection generated by the network itself.

And finally, on strategies toward multi-offenders and co-offenders, the study shows partial but limited adaptation. Multi-offenders were sometimes charged under different statutes, but without systematic use of various combinations. For co-offenders, triangular imputations were recurrent, but prosecutors rarely expanded to multiple and more complex configuration combinations. Consistently employing offense combinations and more robust joint-charging strategies could amplify prosecutorial reach.

\subsection{Limitations and future research}

The criminal network analyzed in this study is reconstructed from publicly available information and investigative reports. As such, it necessarily omits undetected actors and hidden ties that did not surface in formal proceedings. This censoring restricts our ability to capture the full scope of illicit relational structures and likely underestimates the degree of embedded collaboration within the network. Future research could address this limitation by incorporating additional sources, such as indictments, judicial files, and court documentation, which would not only strengthen the evidentiary base but also allow for a more precise differentiation of tie types. Such refinements would enable the application of multiplex or weighted network approaches, thereby offering a more comprehensive account of the complexity and intensity of illicit collaborations.

At the same time, while our ERGM results point to discernible prosecutorial logic, it is important to distinguish substantive findings from potential artifacts of design. Some patterns likely reflect deliberate strategies—such as prioritizing high-severity offenses or targeting central public officials—while others may stem from methodological constraints, including binary charge coding, the aggregation of a decade of prosecutions into one cross-sectional network, and data sparsity. Thus, certain regularities may reflect the modeling framework rather than prosecutorial intent alone.

The cross-sectional design aggregates more than a decade of prosecutions into a static snapshot. While this design enables us to model structural regularities, it obscures temporal dynamics and prosecutorial sequencing. Future research would benefit from employing dynamic models such as Temporal ERGMs (TERGMs) \cite{robins2001} or Stochastic Actor-Oriented Models (SAOMs) \cite{snijders2010, snijders2013}, which can better account for how prosecutorial strategies evolve across time, respond to political shocks, and shape network resilience.

Our empirical scope is constrained to high-profile cases prosecuted by CICIG–FECI. These cases are not fully representative of routine prosecutorial practice in Guatemala and other countries, as they focus on emblematic corruption schemes involving political and economic elites. Expanding the dataset to include more ordinary prosecutions could yield a more representative picture of prosecutorial strategies and their systemic implications.

The findings are also theoretically bounded by the unique institutional context of Guatemala’s special arrangement between CICIG and FECI. While this collaboration enabled innovative prosecutorial strategies, it was simultaneously limited by political interference and judicial inertia \cite{CICIG2019, WOLA2022}. Outcomes in other case studies will depend heavily on the autonomy, impartiality, and institutional capacity of prosecutorial bodies, which vary across countries and over time. As such, generalizing from this case requires caution, particularly in settings without comparable hybrid institutions or international support.

Finally, our analysis does not account for incapacitation effects within larger criminal syndicates—an inherently challenging task, given the opacity and adaptability of such organizations \cite{Hedderman2006, diviak2023}. While conviction rates are often treated as conventional performance indicators, they remain insufficient measures of prosecutorial effectiveness. Punitive outcomes alone rarely succeed in dismantling entrenched criminal networks \cite{Bottoms2004, Hedderman2006}, and under certain conditions, sanctions may even exacerbate offending or produce criminogenic effects \cite{Bouffard2010, ChalfinandMcCrary2017}. Therefore, future research should include punitive outcomes as one dimension of analysis but move beyond them to evaluate prosecutorial effectiveness in terms of network disruption. This means examining the extent to which prosecutions reshape opportunity structures, undermine illicit exchanges, and erode systemic resilience. By adopting this broader perspective, scholars can better assess whether prosecutorial interventions achieve not only case-level convictions but also the systemic weakening of political–economic illicit networks.

\section{Conclusion}

To effectively address criminal networks, prosecutorial strategies must fully integrate intelligence and data analysis \cite{Berlusconi2016}. Measuring and monitoring the structure of imputation networks in real-world settings provides valuable insights into prosecutorial performance. It also facilitates an in-depth examination of prosecutorial strategies, promoting more formalized and integrated standards of decision-taking and advancing the ongoing debate on the effectiveness of prosecutorial outcomes.

However, systemic barriers continue to hinder this integration. Inadequate personnel training, limited institutional resources, and outdated technological capabilities further restrict the adoption of intelligence-driven approaches \cite{Castle2008}. These challenges are compounded by entrenched legal cultures, inflexible internal management practices, and institutional reluctance to share information or collaborate across agencies \cite{Ratcliffe2016}. Additionally, these issues extend to the judiciary, where procedural delays and administrative inefficiencies undermine the effectiveness of imputation strategies. Comprehensive reform is essential to overcome these obstacles and strengthen prosecutorial capacity.

The application of a network modeling framework inspired by social-ecological systems, combined with the multilevel ERGM approach to criminal prosecution represents a significant methodological and conceptual contribution as demonstrated in this study. To the best of our knowledge, this is the first instance that such a comprehensive approach has been applied to the analysis of real-world prosecutorial cases, offering a novel socio-legal perspective. However, this approach faces some challenges as previously discussed.

Overall, our approach highlights the need for prosecutorial strategies to evolve beyond a case-by-case model toward a multi-case, multi-offender imputation framework. Moreover, it offers a benchmark for assessing prosecutorial effectiveness, emphasizing that efforts to dismantle complex criminal networks must be proactive, systemic, and grounded in an understanding of interrelationships across cases. This requires a shift in focus toward investigating how legal decisions reverberate through networks of public and private actors that sustain corruption enterprises. Developing imputation networks to define strategic charges should become a central priority—enabling prosecutors to adopt data-driven methodologies and enhance the systemic impact of their interventions \cite{ Castle2008, Ratcliffe2016, Stemen2022}, as well as prevention with guarantees of non-repetition. 

\section*{Appendix A: Brief history of the CICIG}

The CICIG (\textsl{Comisión Internacional contra la Impunidad en Guatemala}, in Spanish) commenced operations in 2007 after an Agreement between the State of Guatemala and the United Nations \cite{UN2006}, at a time when organized crime and cartels had gained significant control over nine out of the country's twenty-two states \cite{Brands2010}. CICIG identified the Illegal Clandestine Security Apparatuses or CIACS (\textsl{Cuerpos Ilegales y Aparatos Clandestinos de Seguridad}, in Spanish), later known as Illicit Political and Economic Networks or RPEI (\textsl{Redes Político Económicas Ilícitas}, in Spanish), as the root cause of impunity in the country. Within its dismantling strategy, CICIG also targeted politicians and businessmen who created shell companies to massively divert public funds through procurement contracts \cite{CICIG2019}.

CICIG had certain powers to carry out its mandate, which included: requesting statements, documents, and cooperation from any government official or entity; investigating any person, official, or private entity; presenting criminal charges to Guatemala's Public Prosecutor and joining criminal proceedings as a private prosecutor; to report to the relevant administrative authorities the civil servants who committed administrative offenses and to participate as a third party in resulting disciplinary proceedings; and finally, recommend public policies as well as legal and institutional reforms to congress \cite{UN2006}. Based on these powers, CICIG was successful in uncovering more than 70 criminal networks \cite{Hallock2021, HudsonTaylor2010} and participating in 1540 indictments in over 120 cases \cite{CallandHallock2020}. 

After the unilateral termination of the CICIG’s mandate in 2019 by the Guatemalan State, a significant takeover of the government shifted the curse of justice \cite{IACHR2021}. It has been documented that business, political, and military linked to the pre-existing illicit elites uncovered by CICIG took control over the government and the Attorney General's office to target journalists, civil society organization leaders, judges, and public servants who participated in the conviction of individuals in illicit networks. Judges and former prosecutors were forced to flee the country, and many others were jailed under false legal allegations \cite{IACHR2021, WOLA2022, Mattiache2022, WOLA2022}.

\section*{Appendix B: Cases included in the criminal network}

Details concerning the eight cases investigated by CICIG-FECI (from \cite{CICIG2019} and also CICIG's website: \url{https://www.cicig.org/casos-listado/}) and incorporated into the criminal network of our study are outlined here:

\begin{itemize}

    \item The ``La Línea'' case exposed a criminal network that operated both within and beyond state mechanisms, utilizing political influence at the highest levels to forge a parallel structure. This network manipulated the tax administration system, securing substantial illicit profits for all involved, including President Otto Pérez Molina and Vice President Roxana Baldetti. The criminal operation was characterized by a sophisticated scheme of accepting bribes linked to customs fraud . The exposure of "La Línea" led to significant political repercussions: both Pérez Molina and Baldetti resigned from their posts and were subsequently arrested in 2015, several months before their official terms were due to end.

    \item In April 2015, the ``Bufete de Impunida'' case became explicitly connected to the ``La Línea'' case following the initial arrests. Subsequently, members from the criminal network engaged the services of a law firm to surreptitiously secure impunity for several detained individuals. These illicit proceedings were conducted before the court of Judge Marta Sierra de Stalling, and she was formally accused by September 2015 of passive bribery and prevarication, resulting in her being placed under preventive detention.

    \item The case involving Gudy Rivera, an ex-congressman, relates to the judicial nomination process of 2014 in Guatemala. Rivera exerted pressure on Judge Claudia Escobar, promising her re-election as a magistrate in exchange for a favorable ruling on behalf of Vice President Roxana Baldetti. Escobar filed a complaint against this corrupt proposal with the CICIG in 2014. Despite the delayed judicial process, in October 2016, Rivera was convicted and sentenced to 13 years and 4 months in prison for his attempt of bribery \cite{fmm2020}.

    \item The ``State Capture'' case is associated with the Partido Patriota under former President Otto Pérez Molina. Beginning as early as 2008, the party established a criminal network designed to secure resources for electoral campaigns and to facilitate the personal enrichment of its principal members. Several business groups participated in this scheme with the expectation of securing lucrative government contracts and influencing public policy decisions. After assuming office in January 2012, the PP-government and its network strategically occupied key positions within the central government, orchestrating a comprehensive corruption scheme centered on the illicit allocation of state contracts and the systematic collection of kickbacks. Private actors in these arrangements included a monopoly on open television, major telecommunications firms, and various construction companies.

    \item The investigation into the ``La Coperacha'' case revealed that corruption at the highest levels of government under Otto Pérez Molina (2012-2015) was not only tolerated but encouraged. Ministers were expected to establish networks to secure public funds and contribute to gifts offered to high level politicians, leading to widespread practices of embezzlement and kickbacks. Between 2012 and 2014, Vice President Baldetti and several ministers organized a collection of money (``coperacha'', in Spanish) to buy extravagant gifts for President Pérez Molina, such as a boat costing about USD 200,000; a beach house on the Pacific coast payed with the contribution of USD 200,000 by each minister; and a helicopter valued at USD 3.5 million. In 2014, President Pérez Molina also arranged a collection to buy Vice President Baldetti a house in Roatán, Honduras, worth USD 1.2 million. 

    \item In ``Caso TCQ'', during the administration of the Partido Patriota, a corrupt agreement was negotiated between the public port authority and a private firm (TCQ S.A.), involving USD 30 million in bribes for the central government and about USD 3 million for the local government where the port is localized. The contract granted a 25-year renewable lease on 34 hectares of land for the construction and operation of a private container terminal on Guatemala’s southern coast. The case further implicates a network of undue influence, orchestrated to ensure impunity and facilitate the contract's implementation.

    \item The case ``Registro de Información Catastral: caja de pagos'' details a scheme of “phantom jobs” at the Cadastral Information Registry, which were used as a form of payment to fulfill political favors and secure illicit economic benefits. According to the investigators, key beneficiaries of this scheme included the vice president, parliament members, and other high public officers. The investigation revealed financial losses exceeding USD 0.6 million for the institution.

    \item The 2019 case ``Subordination of the Legislative Power to the Executive'' revealed a corruption scheme involving private actors and public officials that compromised Guatemala's legislative independence during the PP-government (2012-2015). The criminal network transformed the Legislative Branch into a facilitator of corrupt interests, affecting key legislative processes, such as the election of congressional boards, the selection of the Comptroller General, and judicial appointments, and favoring private interests, such as the telecommunications company Tigo. This company provided funds used for bribes in cash to each deputy. Executives from Tigo, including Acisclo Valladares Urruela, delivered money to agents who then transferred it to Vice President Roxana Baldetti’s residence and office, where the funds were distributed to representatives. Valladares Urruela later faced charges in the United States for drug trafficking and money laundering linked to the scheme.
    
\end{itemize}

\section*{Appendix C: Laws included in the legal framework}

Brief description of the six laws used in the legal framework network:

\begin{itemize}
    
    \item The Penal Code (Código Penal) covers a broad spectrum of criminal activities, detailing seven specific offenses that range from election-related frauds, such as illicit electoral financing and unregistered electoral financing, to more general crimes including swindle, extortion of public officials, and ideological falsehoods. Additionally, it addresses severe breaches within the judicial system through charges of malfeasance, specifically judicial misconduct, and violations of constitutional laws.

    \item The Law Against Corruption (Ley Contra la Corrupción) is more specialized, focusing on corruption-related offenses. It lists ten specific infractions, including passive and active bribery, embezzlement, and fraud, alongside crimes like influence peddling and illicit enrichment that directly undermine the integrity of public offices. This law also criminalizes actions that obstruct or impede criminal prosecutions, alongside abuse of authority, illegal payments, and breaches of duty, which collectively aim to maintain a high standard of conduct for public officials.

    \item The Law Against Money Laundering (Ley Contra Lavado de Dinero y Otros Activos) singularly targets the laundering of money and other assets, encapsulating the financial crimes associated with disguising illegally obtained funds as legitimate.

    \item The Law Against Customs Fraud (Ley Contra Defraudación y Contrabando Aduanero) deals specifically with offenses related to customs fraud, addressing the evasion of customs duties which is a significant concern for government revenue.

    \item The Law Against Organized Crime (Ley Contra Delincuencia Organizada) focuses on combating organized criminal groups with provisions like the obstruction of justice, intended to penalize efforts that disrupt the legal pursuit of organized crime entities.

    \item The Law Against Drug Trafficking (Ley Contra Narcoactividad) targets criminal associations related to drug trade, underscoring the legal efforts aimed at curbing narcotics-related activities.

\end{itemize}

\small

\bibliographystyle{apalike}
\bibliography{references}

\smallskip
\noindent\textbf{Data availability.} The data used in this article is available from the authors upon request.

\smallskip
\noindent\textbf{Conflict of interests.} The authors declare that the research was conducted in the absence of any commercial or financial relationships that could be construed as a potential conflict of interest.

\smallskip
\noindent\textbf{Authors contributions.} HW and ILP defined the theoretical framework. HW and JRNC curated the data. HW implemented the network models and performed the computational analysis. All authors participated in the development of the conceptual modeling framework, the analysis and discussion of the results, and the writing and approval of the final manuscript.

\smallskip
\noindent\textbf{Acknowledgments.} HW acknowledges support from the European Union’s Horizon 2020 research and innovation programme under the Marie Skłodowska-Curie Grant Agreement No. 859937. JRNC acknowledges support from CONAHCYT's project CF-2019/263958 and grant ``Estancias Posdoctorales por México 2023''. ILP acknowledges the invaluable advice from Jan-Michaele Simon and Rootman Pérez in understanding the prosecutorial strategy of CICIG.


\end{document}